\documentclass[twocolumn,floatfix,pra,showpacs]{revtex4}
\usepackage{graphicx}
\usepackage{amssymb}
\usepackage{amsmath}
\usepackage{color}
\usepackage{psfrag}
\usepackage{epsfig}
\usepackage{bbm}
\usepackage{bm}
\newcommand{\ket}[1]{| #1 \rangle}

\newcommand{\rb}[1]{\left( #1 \right)}

\newcommand{\ew}[1]{\langle #1 \rangle}
\newcommand{\beq}{\begin{eqnarray}}
\newcommand{\eeq}{\end{eqnarray}}

\newcommand{\op}[2]{| #1 \rangle \langle #2 |}

\newcommand{\eq}[1]{Eq.~(\ref{#1})}

\newcommand{\bs}[1]{\boldsymbol{#1}}
\newcommand{\kett}[1]{| #1 \rangle\!\rangle }
\newcommand{\braa}[1]{\langle\!\langle #1|}
\newcommand{\eww}[1]{\langle\! \langle #1\rangle\! \rangle}
\newcommand{\opp}[2]{| #1 \rangle\! \rangle\langle\! \langle #2 |}
\begin{document}
\title{Quantum dynamics in nonequilibrium environments}
\author{Clive Emary}
\affiliation{
  Institut f\"ur Theoretische Physik,
  Hardenbergstr. 36,
  TU Berlin,
  D-10623 Berlin,
  Germany
}

\date{\today}
\begin{abstract}

We present a formalism for studying the behaviour of quantum systems coupled to nonequilibrium environments exhibiting nonGaussian fluctuations.  We discuss the role of a qubit as a detector of the statistics of environmental fluctuations, as well as nonMarkovian effects in both weak and strong coupling limits.  We also discuss the differences between the influences of classical and quantum environments.
As examples of the application of this formalism we study the dephasing and relaxation of a charge qubit coupled to nonequillibrium electron transport through single and double quantum dots.
\end{abstract}
\pacs{03.65.Yz,05.40.-a,05.60.Gg,73.23.-b}
\maketitle

The standard paradigm of system-environment interactions in quantum mechanics employs an equilibrium environment which is large enough that its fluctuations are Gaussian \cite{weiss}.  This model is inappropriate, however, if our quantum system couples strongly to a small number of environmental degrees of freedom, which will typically be out of equilibrium and display a full spectrum of fluctuations.
The best studied example, both in theory\cite{pal02,gal05,gri05,ku05,sch06,abe08} and in  experiment\cite{sim04,ith05,tia07}, is the case of two-level fluctuators in the environment of a Josephson qubit\cite{mak01}.  If the number of fluctuators coupled to the qubit is small,
the decoherence of the qubit shows evidence of nonGaussian environmental fluctuations.
Another important class of such environments is provided by mesoscopic transport, in which a quantum system is influenced by the nonequillibrium transport of electrons through some device.  Examples of such environments include the partition noise from a quantum point contact \cite{ave05,ned07}, and transport through a single-electron transistor (SET) \cite{gur08}.

In this article, we describe a general theory of the influence of nonequillibrium, nonGaussian environments on quantum dynamics.  We work within a generalised master equation (GME) framework \cite{gur98,Brandes04}, and assume that the environmental degrees of freedom to which our quantum system directly couples may be described by a Markovian GME of the Lindblad form.
Under this assumption we derive an effective Liouvillian describing the reduced dynamics of the system alone that can be expressed in terms of environmental correlation functions.  We thus obtain an explicit account of the effects on the system of environmental fluctuations of all orders.

Whilst this theory is presented in general terms, we specifically have in mind applications in mesoscopic transport, where we seek to describe the dephasing and relaxation of a quantum system due to the charge fluctuations of a nearby mesoscopic device.
In Ref.~\cite{gur08} the behaviour of a charge qubit was related to the charge-noise spectrum of SET environment.
It is one of the aims of this work to place the results of Ref.~\cite{gur08} in a broader context and to generalise not only to arbitrary mesoscopic devices in the Coulomb blockade regime, but also to incorporate the effects of charge-fluctuations of orders beyond Gaussian.
For a qubit coupled to the environment via a pure-dephasing coupling, we describe how the long-time behaviour of the qubit is related to the cumulant generating function of the operator through which the system couples to the environment, and show how to calculate this quantity for arbitrary environments.

As illustration of this theory we consider a charge qubit couplied to two mesoscopic environments:
i) the SET environment of Ref.~\cite{gur08}, which is equivalent to a source of classical telegraph noise\cite{ste90}, and, in certain limits, a model of a single background charge fluctuator\cite{pal02,abe08},
and ii) a double quantum dot (DQD) environment. 
Whilst transport through a DQD has been extensively studied \cite{Brandes04,ell02,sto96,EMAB07,kiess07}, to our knowledge, its role as a decoherence source remains unexplored.
Moreover, the DQD environment is an important example because, whereas the SET model can be described in purely classical terms, the inter-dot coherence of the DQD means that transport through it, and hence the fluctuations to which the charge qubit couples, are quantum mechanical in nature.
Both these examples exhibit interesting nonMarkovian qubit dynamics, including dramatic visibility oscillations in the strong coupling limit\cite{ned07}. The DQD model also exhibits a pronounced quantum Zeno effect \cite{Zeno} in this same limit. Comparison of these models highlights the distinctions between quantum and classical fluctuations in determining the dephasing and relaxation of a system coupled to them.

This paper is organised as follows. We first describe the general model considered here and its description in terms of coupled GMEs.  We then show how the environment may be traced out to a yield an effective Liouvillian for the system.
This Liouvillian is related to environmental correlation functions, used to derive dephasing and relaxation rates for the system in the weak coupling limit.  Two special cases are then discussed in which the results are particularly simple: pure dephasing and classical environments with relaxation. We conclude with a study of our two examples and discussions.

\section{System-Environment Model}
\begin{figure}[tb]
  \begin{center}
  \psfrag{ml}{{\huge $\mu_L$}}
  \psfrag{mr}{{\huge $\mu_R$}}
  \epsfig{file=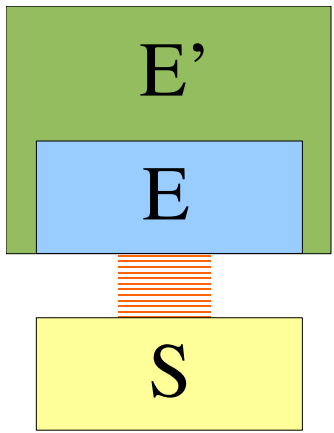, clip=true,width=0.5\linewidth}
  \caption{
    Quantum system S is coupled to an environment divided into two parts: E, which couples directly to S, and E', which does not. E' is large, in equilibrium, and weakly coupled to E such that E is maintained in nonequilibrium.
    \label{FIG_see}
   }
  \end{center}
\end{figure}

Figure \ref{FIG_see} depicts the  general situation under discussion here.  The environmental degrees of freedom are divided into two sets, labelled E and E', according to whether they couple to quantum system S directly or not. 
The Hamiltonian of the system-environment complex is
$
{\cal H} = {\cal H}_\mathrm{S} +  {\cal H}_\mathrm{E} +  {\cal H}_\mathrm{E'} 
  + g{\cal V}_\mathrm{SE} + {\cal V}_\mathrm{EE'}
$,
with ${\cal H}_\mathrm{S,E,E'}$ the isolated Hamiltonians of our decomposition, ${\cal V}_\mathrm{SE}$ and ${\cal V}_\mathrm{EE'}$ interaction Hamiltonians between system and environment E, and between environmental components, and $g$ a dimensionsless coupling constant.
%
%
We assume that the EE' coupling is weak, that reservoir E' is in equilibrium, and that the Born-Markov approximation is valid for the EE' coupling.  Following a standard master equation derivation we trace out environment E' and obtain a GME for the SE density matrix:
\beq
  \partial_t\rho^\mathrm{SE} =
  {\cal L}^\mathrm{SE} \rho^\mathrm{SE}
  =
  \rb{
    {\cal L}^\mathrm{S}_0 + {\cal L}^\mathrm{E}_0 
    + g{\cal M}^\mathrm{SE}
  }\rho^\mathrm{SE}
  ,
\eeq
with system Liouvillian 
$
  {\cal L}^\mathrm{S}_0\rho^\mathrm{S} = -i \left[{\cal H}_\mathrm{S}, \rho^\mathrm{S}\right]
$,
SE coupling Liouvillian
$
  {\cal M}\rho^\mathrm{SE} = -i \left[{\cal V}_\mathrm{SE}, \rho^\mathrm{SE}\right]
$, and Liouvillian ${\cal L}^\mathrm{E}_0$ given by a Lindblad form obtained by tracing out E'.

In the following we will employ a notation for GMEs in which the elements of the density matrix are arranged into a vector $\kett{\rho^\mathrm{SE}}$ with populations first, followed by coherences \cite{jah05}.  In this notation, superoperators are written as matrices, and the GME for the SE density matrix, now a vector, has the form
\beq
  \partial_t \kett{\rho^\mathrm{SE}} &=& 
  {\cal L}^\mathrm{SE} \kett{ \rho^\mathrm{SE} }
  =
  \rb{
    {\cal L}_0^\mathrm{S}+ {\cal L}_0^\mathrm{E} +g {\cal M}
  }\kett{\rho^\mathrm{SE}}
  \nonumber\\
  &=&
  \rb{
    {\cal L}_0^\mathrm{SE} +g {\cal M}
  }\kett{\rho^\mathrm{SE}}
  \label{GMESE}
  .
\eeq
The GME for the environment reads
$  
  \partial_t \kett{\rho^\mathrm{E}} =
  {\cal L}_0^\mathrm{E}\kett{\rho^\mathrm{E}}
$.
Let us denote the eigenvalues of ${\cal L}_0^\mathrm{E}$ as $\lambda_n^\mathrm{E}$, assumed distinct, and its right and left eigenvectors  as $\kett{\phi^\mathrm{E}_n}$ and $\braa{\phi^\mathrm{E}_n}$ respectively.
These vectors form a biorthogonal set, $\eww{\phi^\mathrm{E}_m | \phi^\mathrm{E}_n} = \delta_{nm}$, but are not adjoint, since ${\cal L}_0^\mathrm{E}$ is nonHermitian.
The stationary state of the environment is given by $\kett{\rho_\mathrm{stat}^\mathrm{E}}=\kett{\phi^\mathrm{E}_0}$, the zero-eigenvalue eigenvector of ${\cal L}_0^\mathrm{E}$, i.e. 
${\cal L}_0^\mathrm{E}\kett{\phi^\mathrm{E}_0}=\lambda_0^\mathrm{E}\kett{\phi^\mathrm{E}_0} =0$.
The corresponding left-eigenvector $\braa{\phi^\mathrm{E}_0}$, has elements 1 at locations corresponding to populations and is zero otherwise.  Similar definitions hold for the free system Liouvillian $ {\cal L}_0^\mathrm{S}$ and its eigendecomposition; but note that, since $ {\cal L}_0^\mathrm{S}$ contains no damping, its nullspace will be of dimension greater than one.

Finally we assume that the SE interaction has the bilinear form,
$
  {\cal V}_\mathrm{SE} = \frac{1}{2} \sigma \epsilon
$,
where $\sigma$ is a dimensionless system operator and $\epsilon$ an environment operator with dimensions of energy.  The  corresponding Liouvillian is obtained from
$
  {\cal M}\rho^\mathrm{SE} =
  -i \frac{1}{2} \rb{\sigma \epsilon\rho^\mathrm{SE} -\rho^\mathrm{SE} \sigma \epsilon}
$,
which we write in vector notation as
\beq
  {\cal M}\kett{\rho^\mathrm{SE}} &=&
  -i  \frac{1}{2} \rb{O^+_\sigma O^+_\epsilon - O^-_\sigma O^-_\epsilon}\kett{\rho^\mathrm{SE}}
  .
  \label{Mvecnot}
\eeq
The superoperators $O^\pm_\epsilon$, here represented as matrices, can be obtained by considering matrix elements.


Although we will derive results for arbitrary systems, it is often useful to discuss the case when the system is a qubit. In its diagonal basis, the qubit Hamiltonian is $H_S = \frac{1}{2}\Delta \sigma_z$ and, in the basis $\rb{\rho_{11},\rho_{22},\rho_{12},\rho_{21}}$, the corresponding free Liouvillian is 
\beq
  {\cal L}^\mathrm{S}_0 &=&
  \rb{
  \begin{array}{cccc}
    0 & 0 & 0 & 0 \\
    0 & 0 & 0 & 0 \\
    0 & 0 & i\Delta & 0 \\
    0 & 0 & 0 & -i \Delta
  \end{array}
  }
  \label{QBL0}
\eeq
The system part of the SE coupling operator $\sigma$ is then a traceless Hermitian matrix with elements $\mathbf{n}\cdot\bs{\sigma}$ where $\mathbf{n}$ is a unit vector and $\bs{\sigma}$ the vector of Pauli matrices.  The relevant operators in Liouville space are
\beq
  O_\sigma^+ &=&
  \rb{
  \begin{array}{cccc}
    n_z& 0 & 0 & n_- \\
    0 & -n_z & n_+ & 0 \\
    0 & n_- & n_z & 0 \\
    n_+ & 0 & 0 & -n_z
  \end{array}
  }
  \nonumber\\
  O_\sigma^- &=&
  \rb{
  \begin{array}{cccc}
    n_z& 0 &  n_+ & 0\\
    0 & -n_z & 0 & n_- \\
    n_- & 0 & -n_z & 0 \\
    0 & n_+ & 0 & n_z
  \end{array}
  }
  \label{QBmatrices}
\eeq
with $n_\pm = n_x \pm n_y$. We will discuss two particular examples in the following:

\noindent {\it Pure dephasing:} With $\sigma=\sigma_z$, the coupling is diagonal in the same basis as the free evolution of the system.  In this case, we have a `pure dephasing' model with both $O_\sigma^\pm$ diagonal:
\beq
  O_\sigma^+ =
  \rb{
  \begin{array}{cccc}
    1& 0 & 0 & 0 \\
    0 & -1 & 0 & 0 \\
    0 & 0 & 1 & 0 \\
    0 & 0 & 0 & -1
  \end{array}
  }
  ;\quad
  O_\sigma^- =
  \rb{
  \begin{array}{cccc}
    1& 0 & 0 & 0 \\
    0 & -1 & 0 & 0 \\
    0 & 0 & -1 & 0 \\
    0 & 0 & 0 & 1
  \end{array}
  }
  .
\eeq
Under this coupling, only the off-diagonal elements of the qubit density matrix evolve in time, and we write $\rho_{01}(t)= D(t)\rho_{01}(0)$ with the `degree of coherence' $D(t)$ to be determined.

~\\
\noindent {\it Orthogonal coupling:} The other coupling that we will explicitly consider is `orthogonal coupling', with  $\sigma=\sigma_x$.  In this case we have the off-diagonal super-operator matrices
\beq
  O_\sigma^+ =
  \rb{
  \begin{array}{cccc}
    0 & 0 & 0 & 1 \\
    0 & 0 & 1 & 0 \\
    0 & 1 & 0 & 0 \\
    1 & 0 & 0 & 0
  \end{array}
  }
  ;\quad
  O_\sigma^- =
  \rb{
  \begin{array}{cccc}
    0 & 0 & 1 & 0 \\
    0 & 0 & 0 & 1 \\
    1 & 0 & 0 & 0 \\
    0 & 1 & 0 & 0
  \end{array}
  }
  \label{OEOG}
  ,
\eeq
and this coupling will induce relaxation in the system.

Finally, we will also refer to the situation when the environment is {\it classical}.  In this case, only populations are required to describe the state of environment E, and the GME determining  its behaviour is actually a rate equation.  It then follows that operators acting on E are diagonal and commute at different times.  This in turn implies that the $\pm$ superoperators are identical: $O^-_\epsilon = O^+_\epsilon$.  As will be made clear below, this situation represents an environment which experiences no back-action due to its interaction with the system.

\section{Effective system Liouvillian}
Having set up our model of coupled master equations, we now proceed to derive a description of the system's behaviour in terms of environmental quantities.  To this end, we derive an effective Liouvillian for the system dynamics.  
Laplace transform of \eq{GMESE} gives 
$
  \kett{\rho^\mathrm{SE}(z)} =  \Omega^\mathrm{SE}(z)
  \kett{\rho^\mathrm{SE}(t=0)}
$ with SE propagator 
$ 
  \Omega^\mathrm{SE}(z) = \left[z- {\cal L}^\mathrm{SE} \right]^{-1}
$.
This we expand in orders of $g$ as
\beq
  \Omega^\mathrm{SE}(z) 
  &=& \Omega^\mathrm{SE}_0(z) +  g\Omega^\mathrm{SE}_0(z) {\cal M} \Omega^\mathrm{SE}_0(z)
  \nonumber\\
  &&
  +  g^2\Omega^\mathrm{SE}_0(z) {\cal M} \Omega^\mathrm{SE}_0(z) {\cal M} \Omega^\mathrm{SE}_0(z)
  +\ldots
  \label{Omexp}
  ,
\eeq
with
$
  \Omega^\mathrm{SE}_0(z) = \left[z- {\cal L}_0^\mathrm{SE} \right]^{-1}
$,
the free SE propagator.
Assuming that the environment starts in its steady-state $\rho_\mathrm{stat}^\mathrm{E}$, the reduced system propagator $\Omega^\mathrm{S}(z)$ is given by
$
  \Omega^\mathrm{S}(z) 
  = \braa{\phi^\mathrm{E}_0} \Omega^\mathrm{SE}(z)  \kett{\phi^\mathrm{E}_0} 
$,
corresponding to a trace over the remaining environmental degrees of freedom.
With the expansion of \eq{Omexp}, we have
\beq
  \Omega^\mathrm{S}(z) 
  &=&
  \braa{\phi^\mathrm{E}_0} \Omega^{SE}(z)  \kett{\phi^\mathrm{E}_0} 
  \nonumber\\
  &=& 
  \Omega^\mathrm{S}_0(z)
  +\Omega^\mathrm{S}_0(z) \braa{\phi^\mathrm{E}_0} {\cal M} \kett{\phi^\mathrm{E}_0} \Omega^\mathrm{S}_0(z)
  \nonumber\\
  &&
  +\Omega^\mathrm{S}_0(z) \braa{\phi^\mathrm{E}_0} {\cal M} \Omega^{SE}_0(z) {\cal M} 
    \kett{\phi^\mathrm{E}_0} \Omega^\mathrm{S}_0(z)
  +\ldots  
  \label{OmS1}
  ,
\eeq
where we have identified the free system propagator
\beq
  \Omega^\mathrm{S}_0(z) 
  &=&  \braa{\phi^\mathrm{E}_0} \Omega^{SE}(z)  \kett{\phi^\mathrm{E}_0} 
  = \frac{1}{z- {\cal L}_0^\mathrm{S}}
  .
\eeq
We can consider the system propagator as arising from an effective system Liouvillian, 
${\cal L}_\mathrm{eff}^\mathrm{S}(z)$, which will be nonMarkovian. We write this as a series in $g$:
\beq
  {\cal L}^\mathrm{S}_\mathrm{eff}(z) = \sum_{n=0}^\infty g^n {\cal L}_n^\mathrm{S} (z)
  ,
\eeq
such that the full system propagator can also be written as 
\beq
  \Omega^\mathrm{S}(z) &=& \frac{1}{z-{\cal L}_\mathrm{eff}^\mathrm{S}}
  \nonumber\\
  &=&
  \Omega^\mathrm{S}_0(z)
  + \Omega^\mathrm{S}_0(z){\cal L}_1^\mathrm{S}  \Omega^\mathrm{S}_0(z)
    + \Omega^\mathrm{S}_0(z) {\cal L}_2^\mathrm{S}  \Omega^\mathrm{S}_0(z)
  \nonumber\\
  &&
    + \Omega^\mathrm{S}_0(z){\cal L}_1^\mathrm{S}  \Omega^\mathrm{S}_0(z) {\cal L}_1^\mathrm{S}  \Omega^\mathrm{S}_0(z)
    +\ldots
  \label{OmS2}
  .
\eeq
An order-by-order comparison of \eq{OmS1} and \eq{OmS2} gives us
\beq
  {\cal L}_n^\mathrm{S} 
  =
  \braa{\phi^\mathrm{E}_0}
   \left\{ {\cal M} 
     \Omega^{SE}_0(z) {\cal Q}^\mathrm{E} 
   \right\}^{n-1}
   {\cal M}
  \kett{\phi^\mathrm{E}_0}
  ;~ n \ge 1.
  \label{Ln}
\eeq
Formal resummation yields
\beq
 {\cal L}^\mathrm{S}_\mathrm{eff}(z)
  = 
  {\cal L}_0^\mathrm{S}
  +g
   \braa{\phi^\mathrm{E}_0}
   \frac{1}
   {
     \mathbbm{1}^{SE }
     -
    g{\cal M} \Omega^\mathrm{SE}_0(z) {\cal Q}^\mathrm{E} 
   }
   {\cal M}
  \kett{\phi^\mathrm{E}_0}
  \label{LeffFULL}
  ,
\eeq
where $ {\cal Q}^\mathrm{E} =  \mathbbm{1}^\mathrm{E} -\opp{\phi^\mathrm{E}_0}{\phi^\mathrm{E}_0}$, the projector out of the environment steady state. Using the eigendecomposition of the free SE Liouvillian, we can write the free SE propagator as
\beq
  \Omega_0^{SE}(z) 
  &=& 
  \frac{1}{z - {\cal L}_0^\mathrm{S} + {\cal L}_0^\mathrm{E}}
  \nonumber\\
  &=&
  \sum_{n,\nu=0}
  \frac{1}{z - \lambda_\nu^\mathrm{S} -\lambda_n^\mathrm{E}}
  \kett{\phi^\mathrm{S}_\nu \phi^\mathrm{E}_n}\braa{\phi^\mathrm{S}_\nu \phi^\mathrm{E}_n}
  \nonumber\\
  &=&
   \sum_{\nu=0} \kett{\phi^\mathrm{S}_\nu}\braa{\phi^\mathrm{S}_\nu}
   \otimes
   \Omega_0^\mathrm{E}(z_\nu)
   ,
\eeq
with $z_\nu \equiv z - \lambda_\nu^\mathrm{S}$.  The effective Liouvillian may then be written
\beq
  {\cal L}^\mathrm{S}_\mathrm{eff}(z)
  &=& 
  {\cal L}^\mathrm{S}_0 
  \nonumber\\
  &&
  \!\!\!\!\!\!\!\!\!\!\!\!  \!\!\!\!\!\!\!\!
  +
   g\braa{\phi^\mathrm{E}_0}
   \frac{1}
   {
     \mathbbm{1}^{SE }
     -  
    g \sum_\nu
     {\cal M}
     \opp{\phi_\nu^\mathrm{S}}{\phi_\nu^\mathrm{S}}
     \Omega^\mathrm{E}_0(z_\nu){\cal Q}^\mathrm{E}
   }
   {\cal M}
  \kett{\phi^\mathrm{E}_0}.
  \nonumber\\
  \label{LeffALT}
\eeq

The effective Liouvillian of \eq{LeffFULL} or \eq{LeffALT} is the main formal results of this work; it describes the system dynamics in a compact, self-contained form and includes environmental fluctuations of all orders. In this form it is not particularly instructive, however, since both S and E quantities appear in an intertwined way.  In order to see the significance of these results then, we will consider first a weak coupling expansion, and then some special cases where S and E dependencies can be separated.

Before doing so, let us note that an expression similar to \eq{OmS1} can be written down for the reduced environmental propagator, $\Omega^\mathrm{E}(z)$. 
The initial state of our system, which we take to be a qubit here, is arbitrary and can be written in the form
\beq
  \kett{\rho^\mathrm{S}_0} &=& \kett{\phi^\mathrm{S}_0} 
  + \frac{a}{2} \kett{\phi^\mathrm{S}_1}
  + \frac{b}{4}\kett{\phi^\mathrm{S}_2}
  + \frac{b^*}{4} \kett{\phi^\mathrm{S}_3}
  \nonumber\\
  &=&
  \frac{1}{2}
  \rb{
    \begin{array}{c}
     1\\ 1 \\0 \\0
    \end{array}
  }
  +
  \frac{a}{2}
  \rb{
    \begin{array}{c}
      -1\\1 \\0 \\0
    \end{array}
  }
  +
  \frac{b}{4}
  \rb{
    \begin{array}{c}
     0\\0 \\1\\0
    \end{array}
  }
  +
  \frac{b^*}{4}
  \rb{
    \begin{array}{c}
      0 \\0 \\0\\1
    \end{array}
  }
  ,
  \nonumber\\
\eeq
which defines the vectors $\kett{\phi_i^\mathrm{S}}$, and coefficients
$-1 \le a \le 1$ and $b\le1-a^2$.  The only conjugate state we shall need here is $\braa{\phi_0^\mathrm{S}} = \rb{1,1,0,0}$.  With the system starting in this state, the environmental propagator can be written as
\beq
  \Omega^\mathrm{E}(z) 
  &=& 
  \eww{\phi_0^\mathrm{S}|\Omega^\mathrm{SE}(z) |\rho_0^\mathrm{S}}
  \nonumber\\
  &=&
  \eww{\phi_0^\mathrm{S}|
  \Omega_0^\mathrm{SE}(z)  + g   \Omega_0^\mathrm{SE}(z){\cal M}\Omega_0^\mathrm{SE}(z)
  +\ldots
  |\rho_0^\mathrm{S}}
  \nonumber\\
  &=&
  \Omega_0^\mathrm{E}(z)
  + g \Omega_0^\mathrm{E}(z)
  \eww{\phi_0^\mathrm{S}|{\cal M}\Omega_0^\mathrm{SE}(z)
  |\rho_0^\mathrm{S}}
  +\ldots
  .
  \nonumber\\
  \label{OMeffE}
\eeq
With this expression we can calculate the effects of back-action of the system on the environment.  We will not follow this calculation further, except to note what happens for classical environments.  In this case, we have $O^-_\epsilon = O^+_\epsilon$ and thus ${\cal M} = i O^+_\epsilon \rb{O^+_\sigma-O^-_\sigma}$.  The effective environment propagator thus contains terms like $ i O^+_\epsilon\eww{\phi_0^\mathrm{S}|\rb{O^+_\sigma-O^-_\sigma}\Omega_0^\mathrm{SE}(z)| \rho_0^\mathrm{S}} $ and, as is easy to verify, $\braa{\phi_0^\mathrm{S}}\rb{O^+_\sigma-O^-_\sigma} =0$.  
All terms beyond the first in \eq{OMeffE} start in just this fashion, and therefore, for classical environments we have $\Omega^\mathrm{E}(z) =\Omega_0^\mathrm{E}(z)$, and there is no back-action.  It is possible to construct models with a classical environment that do experience back-action with, for example, the rates of the free environmental Liouvillian depending on the state of the qubit \cite{mak01,gur08,oxt06}
Such models, however, are outside the class discussed here in which all back-action effects arise from the quantum-mechanical nature of the system-environment coupling.

\section{Weak coupling: dephasing and relaxation rates}

We now consider the situation where the SE coupling is small $g\ll 1$ and describe the weak coupling expansion of \eq{LeffALT}. In this case, we can consider the partial Liouvillians of \eq{Ln} as successive approximations to the full Liouvillian.  At first order, we have
$  
  {\cal L}_1^\mathrm{S} 
  = \textstyle{ \frac{1}{2i} }\ew{\epsilon}\rb{O^+_\sigma - O^-_\sigma}
$,
with $\ew{\epsilon}= \eww{O^+_\epsilon}= \eww{O^-_\epsilon}$ the steady-state expectation value of operator $\epsilon$.

At second order we have
\beq
  {\cal L}_2^\mathrm{S} 
  &=&
  \braa{\phi^\mathrm{E}_0}
   {\cal M}{\cal Q}^\mathrm{E}
   \Omega^{SE}_0(z) {\cal Q}^\mathrm{E} {\cal M}
  \kett{\phi^\mathrm{E}_0}
  \nonumber\\
  &=&
  \sum_{\nu=0}
  \braa{\phi^\mathrm{E}_0}
   {\cal M}
   \opp{\phi_\nu^\mathrm{S}}{\phi_\nu^\mathrm{S}}
   {\cal Q}^\mathrm{E}\Omega^\mathrm{E}_0(z_\nu) {\cal Q}^\mathrm{E} {\cal M}
  \kett{\phi^\mathrm{E}_0}
  .
\eeq
From the form of ${\cal M}$, we see that this expression depends on environmental quantities like 
$ \eww{ O^+_\epsilon {\cal Q}^\mathrm{E} \Omega_0^\mathrm{E}(z_\nu) {\cal Q}^\mathrm{E} O^+_\epsilon}$, which can be evaluated straightforwardly for any particular model. Moreover, they can be related to correlation functions of operator $\epsilon$ via the quantum regression theorem (QRT) \cite{lax68,carBOOK}.
Let us define the second-order correlation function
\beq
  \bar{S}^{(2)}(z_\nu)
  \equiv 
  \int_0^\infty d \tau e^{- z_\nu \tau} \ew{\delta \epsilon(\tau) \delta \epsilon(0)}
\eeq 
with  $\delta \epsilon(t) = \epsilon(t) - \ew{\epsilon} $.  Here the time-dependence of the operators is given by the evolution of the full environmental Hamiltonian
$  
  {\cal H}_\mathrm{E} +  {\cal H}_\mathrm{E'} + {\cal V}_\mathrm{EE'}
$.  
Using the QRT to express this correlation function in terms of quantities acting on E alone, we obtain
$ \bar{S}^{(2)}(z_\nu) = 
  \eww{ O^+_\epsilon \Omega_0^\mathrm{E}(z_\nu)  {\cal Q}^\mathrm{E} O^+_\epsilon}
$.
Similarly, by recalling that super-operator $O^-_\epsilon$ is equivalent to operator $\epsilon$ acting from the right, we obtain
\beq
  \eww{ O^\pm_\epsilon \Omega_0^\mathrm{E}(z_\nu)  {\cal Q}^\mathrm{E} O^+_\epsilon}
  &=&
  \bar{S}^{(2)}(z_\nu)
  \nonumber\\
  \eww{ O^\pm_\epsilon \Omega_0^\mathrm{E}(z_\nu)  {\cal Q}^\mathrm{E} O^-_\epsilon}
  &=&
  \rb{\bar{S}^{(2)}(z^*_\nu)}^*
  .
\eeq
Putting these results together, we obtain our final form for the second-order effective Liouvillian
\beq
  {\cal L}^\mathrm{S}_2(z) &=& (-i g/2 )^2 \sum_{\nu=0}
   \rb{O^+_\sigma - O^-_\sigma}
   \kett{\phi^\mathrm{S}_\nu}\braa{\phi^\mathrm{S}_\nu}
   \nonumber\\
   &&
  \times
   \left\{
     O^+_\sigma \bar{S}^{(2)}(z_\nu)- O^-_\sigma \rb{\bar{S}^{(2)}(z^*_\nu)}^*
   \right\}
   .
   \label{LS2}
\eeq
This Liouvillian determines the system behaviour for all times in the weakly coupled limit for arbitrary system and environment. Its form is simply that of a matrix in system-space, the elements of which contain environmental correlation functions evaluated at various frequencies.

The long-time behaviour of a qubit can be described by a pair of rates, $\Gamma_d$ and $\Gamma_r$ describing dephasing and relaxation respectively.  These rates are determined from 
$ {\cal L}^\mathrm{S}_\mathrm{eff}(z)$, or as is the case here, its second-order approximation.  We first diagonalise  
$ 
  {\cal L}^\mathrm{S}_\mathrm{eff}(z)
  =
  V(z) \Lambda(z) V^{-1}(z)
$, such that the effective system propagator may be written
$
  \Omega^\mathrm{S} =  
  V(z) \frac{1}{z-\Lambda(z)} V^{-1}(z)
$.  We then find the poles of each $\frac{1}{z-\Lambda_{kk}(z)}$, with $\Lambda_{kk}(z)$ the $k$th eigenvalue of $ {\cal L}^\mathrm{S}_\mathrm{eff}(z)$.  In the long time limit, only the pole lying rightmost in the complex plane contributes.  For a qubit then, $\frac{1}{z-\Lambda(z)}$ assumes the asymptotic form
\beq
  \frac{1}{z-\Lambda(z)} \to
  \rb{
    \begin{array}{cccc}
      \frac{1}{z} & 0 & 0 & 0 \\
      0 &  \frac{c_1}{z+ \Gamma_r}& 0 & 0 \\
      0 & 0 &  \frac{c_2}{z+i\nu + \Gamma_d} & 0 \\
      0 & 0 & 0 &   \frac{c_2^*}{z-i\nu + \Gamma_d}
    \end{array}
  }
\eeq
with $c_i$ constants, $\nu$ some frequency of coherent oscillation and $\Gamma_d$ and $\Gamma_r$, the aforementioned rates.

We now discuss \eq{LS2} and the corresponding rates for two illustrative couplings.

\noindent{\it Pure dephasing:}
Let us consider a qubit and set $\ew{\epsilon} = 0$, since we can always incorporate it into a redefinition of $\Delta$. The effective Liouvillian of \eq{LS2} is diagonal, has zeroes at first and second diagonal elements, and has the third element
\beq
  \rb{
    {\cal L}_\mathrm{eff}^\mathrm{S}
  }_{33}
  = l(z)=
  i \Delta -\frac{1}{2} g^2\left\{\bar{S}(z-i \Delta) + \bar{S}^*(-z +i \Delta)\right\} 
  \nonumber\\
  ,
\eeq
and $\rb{{\cal L}_\mathrm{eff}}_{44}=\rb{{\cal L}_\mathrm{eff}}_{33}^*$.  The off-diagonal elements of the qubit density matrix there evolve as $\rho_{01}(z)= D(z)\rho_{01}(t=0)$ with $D(z) = \left\{z-l(z)\right\}^{-1}$.  The rightmost-lying pole of $D(z)$ is
\beq
  z_0 &=&  i \Delta - \frac{1}{2} g^2\left\{\bar{S}(0) + \bar{S}^*(0)\right\}
  ,
\eeq
correct to second order in $g$.  The dephasing rate is therefore 
\beq
  \Gamma_d = \frac{1}{2} g^2 S_\epsilon^{(2)}(0)
  ,
  \label{G2}
\eeq
with the full, symmetrised correlation operator
\beq
  S^{(2)}_\epsilon(\omega)
  = \frac{1}{2}
  \int_{-\infty}^\infty d \tau e^{i\omega \tau} 
  \ew{
    \left\{
      \delta \epsilon(\tau), \delta \epsilon(0)
    \right\}
    }
    \label{S2}
    ,
\eeq
with $\left\{\ldots,\ldots\right\}$ denoting the anticommutator.

\noindent{\it Orthogonal coupling:}  We now consider the orthogonal coupling and set $\ew{\epsilon} = 0$ for simplicity's sake.  The non-zero poles of the second-order effective Liouvillian with system operators of \eq{OEOG} are
\beq
  z_1 &=& -(g/2)^2 
  \left\{
    \bar{S}(-i\Delta) + \bar{S}(i\Delta)
    +\bar{S}^*(-i\Delta)+\bar{S}^*(i\Delta)
  \right\} 
   \nonumber\\
    z_2 &=& 
    i \Delta - (g/2)^2 \left\{ \bar{S}(i \Delta) +  \bar{S}^*(-i \Delta)\right\} 
    \nonumber\\
    z_3 &=& 
    - i \Delta - (g/2)^2 \left\{ \bar{S}(- i \Delta) +  \bar{S}^*( i \Delta)\right\}
    .
\eeq 
The corresponding relaxation and dephasing rates are
\beq
  \Gamma_r &=& z_1 = \frac{1}{2} g^2S_\epsilon^{(2)}(\Delta)
  ;\quad
   \Gamma_d = \mathrm{Re}\rb{z_2} =\frac{1}{2}  \Gamma_r
   ,
\eeq
correct to second order in $g$.
The stationary state of the qubit is found to be
\beq
  \rho^\mathrm{S}_\mathrm{stat} 
  = 
  \rb{
    \begin{array}{cc}
     \frac{\bar{S}(i \Delta) +  \bar{S}^*(i \Delta)}{S^{(2)}_\epsilon(\Delta)} & 0 \\ 
     0 &  \frac{ \bar{S}(-i \Delta) +  \bar{S}^*(-i \Delta) }{S^{(2)}_\epsilon(\Delta)}
    \end{array}
  }
  \label{rhostatOG}
  .
\eeq
We define 
\beq
  P = 2 \mathrm{Tr} \left\{\rho^2 \right\} -1
\eeq
as a measure of the purity of a qubit density matrix.  For a completely mixed state $P=0$ and pure state $P=1$.
For a pure dephasing model, the final purity depends on the initial state --- the generic final state is $\rho =\mathrm{Diag}\rb{c,1-c}$, where $c$ is given by the initial conditions.  The purity of this state is $ P = \rb{2c-1}^2$, which is unity if we start in a pure localised state, $c=0,1$, and zero if we start in the superposition $c=1/2$.  For the stationary state of the orthogonal coupling model, \eq{rhostatOG}, the purity is
\beq
  P = 
  \rb{
  \frac{1}{S^{(2)}_\epsilon(\Delta)}
  \int_{-\infty}^\infty d \tau e^{i\omega \tau} 
  \ew{
    \left[
      \delta \epsilon(\tau), \delta \epsilon(0)
    \right]
    }
  }^2
\eeq
with $\left[\ldots , \ldots\right]$, the commutator.  In words: the purity is determined by the ratio of the Fourier transforms of the commutator and the anticommutator of fluctuation-operator $\delta\epsilon$ at different times.  For a classical environment, the commutator is zero, and the purity is zero.  Nonzero values of the purity are an indicator of the existence of back-action of the system on the environment\cite{gur08}, and in the current class of models, the back-action is always quantum.

These results illustrate the generality of the connexion between these rates and the second-order correlation functions given in Ref.~\cite{gur08}.
In principle, we can extend the above analysis to arbitrary order in $g$, making the connexion to environmental correlations functions with the QRT.  At third order, for example, we have the
effective Liouvillian 
\beq
  {\cal L}_3^\mathrm{S} 
  &=&
  \braa{\phi^\mathrm{E}_0}
   {\cal M}{\cal Q}^\mathrm{E}
   \Omega^{SE}_0(z) {\cal Q}^\mathrm{E} {\cal M}{\cal Q}^\mathrm{E}
   \Omega^{SE}_0(z) {\cal Q}^\mathrm{E} {\cal M}
  \kett{\phi^\mathrm{E}_0}
  .
  \nonumber\\
\eeq
With third-order correlation functions defined as in appendix \ref{appO3},
this Liouvillian can be written as
\begin{widetext}
\beq
 {\cal L}^\mathrm{S}_3 &=& (-i g/2 )^3 \sum_{\nu,\nu'=0}
   \rb{O^+_\sigma - O^-_\sigma} 
   \kett{\phi^\mathrm{S}_\nu}\braa{\phi^\mathrm{S}_\nu}
   \left\{
    O^+_\sigma \kett{\phi^\mathrm{S}_{\nu'}}\braa{\phi^\mathrm{S}_{\nu'}}
    \rb{
      O^+_\sigma \bar{S}_\epsilon^{(3a)}(z_{\nu}, z_{\nu'})  
      -
      O^-_\sigma  \bar{S}_\epsilon^{(3b)}(z_{\nu'}, z_{\nu})
    }
   \right.
   \nonumber\\
   &&
   ~~~~~~~~~~~~~~~~~~~~~~~~~~~~~~~~~~~~~~~~~
   \left.
   -
   O^-_\sigma \kett{\phi^\mathrm{S}_{\nu'}}\braa{\phi^\mathrm{S}_{\nu'}}
    \rb{
      O^+_\sigma \rb{\bar{S}_\epsilon^{(3b)}(z^*_{\nu'}, z^*_{\nu})  }^*
      -
      O^-_\sigma  \rb{ \bar{S}_\epsilon^{(3a)}(z^*_{\nu}, z^*_{\nu'})  }^*
    }
   \right\}
   \label{L3}
   .
\eeq
\end{widetext}

\section{Pure Dephasing \label{SECPD}}

The Liouvillian of \eq{LeffFULL} contains a mixture of system and environment operators in the inverse and in general this means that this inverse can not be carried out explicitly.  Useful results can be obtained by expansion, as above, but the results can becomes unwieldy.
In certain cases, however, we can effect a separation of S and E components, expressing the matrix elements of ${\cal L}^\mathrm{S}_\mathrm{eff}$ in terms of environmental quantities.  In this section and the next, we consider two cases: pure dephasing and orthogonal coupling to a classical environment.

In the pure dephasing model, the effective Liouvillian is diagonal, and only has non-zero elements at positions 3 and 4.
The degree of coherence $D(z)$ is given by the third diagonal element of $\Omega^\mathrm{S}(z)$.  In an interaction picture for the system (obtained by shifting $z\to z-{\cal L}_0^\mathrm{S}$), we obtain
\beq
  D(z) = \left[z - l(z) \right]^{-1}
\eeq
with
\beq
  l(z)=
  -i g
  \eww{ \phi^\mathrm{E}_0|
    \left[
      \mathbbm{1}^\mathrm{E} + i g  O_\epsilon \Omega^\mathrm{E}(z){\cal Q}
    \right]^{-1}
    O_\epsilon
  |\phi^\mathrm{E}_0}
  \label{LPD}
  ,
\eeq
where we have used the shorthand $ O_\epsilon = \frac{1}{2}\rb{O^+_\epsilon + O^-_\epsilon}$. 
This quantity
contains information on environmental fluctuations of all orders,
as may be seen by expansion.  However, expansion is not necessary because, for a finite-dimensional environment E at least, it may be evaluated directly in closed form through matrix inversion.  Evaluation for an infinite-dimensional environment may still be possible using phase-space methods \cite{carBOOK}.

Equation \eq{LPD} also suggests the point-of-view of the pure-dephasing qubit acting as a detector of the environmental fluctuations.  As we now show, $D(t)$, the inverse Laplace transform of $D(z)$, is related to the generating function for the zero-frequency cumulants of operator $\epsilon$.
The inverse Laplace transform of $D(z)$ is obtained from the roots of 
$
  z - l(z) =0
$ \cite{fli08}.  In the long time limit, we only need the pole lying rightmost in the complex plane, which we denote $\tilde{z}$, such that we can approximate $D(t) = c e^{\tilde{z} t}$ with $c$ some constant.  Considering $g$ as infinitesimally small, we can write $\tilde{z}$ as an expansion about $z=0$: $\tilde{z} = \sum_{n=1} (-i g)^n \tilde{z}_n$ and solve
 $
  \tilde{z} - l(\tilde{z}) =0
$ order-by-order. For example, the first three terms are
\beq
  \tilde{z}_1 &=& \eww{O_\epsilon }
  ,~~~
  \tilde{z}_2 = \eww{O_\epsilon \Omega^\mathrm{E}(0){\cal Q} O_\epsilon}
  ,
  \nonumber\\
   \tilde{z}_3 &=& 
   \eww{
   O_\epsilon \Omega^\mathrm{E}(0){\cal Q} 
   \rb{O_\epsilon - \eww{O_\epsilon}} \Omega^\mathrm{E}(0){\cal Q} 
   O_\epsilon 
   }
   ,
   \label{ztilcomp}
\eeq
with the shorthand $\eww{\ldots} = \eww{\phi_0^\mathrm{E}|\ldots |\phi_0^\mathrm{E}}$.
Application of the QRT, shows that these quantities are equal to the zero-frequency limits of the Keldysh-ordered correlation functions: $\tilde{z}_1 = \ew{\epsilon}$, $\tilde{z}_2$ is equal to the zero-frequency limit of $S^{(2)}_\epsilon (\omega)$ from \eq{S2}, and $\tilde{z}_3$ is the zero-frequency limit of $S^{(3)}_\epsilon (\omega,\omega')$ defined in appendix \ref{appO3}, and so on.  Therefore, within the approximations made here, $\tilde{z} $ may be identified as the cumulant generating function for zero-frequency, Keldysh-ordered correlation functions of the operator, which we denote  $\delta\epsilon$, $S_\epsilon^{(n)} (\left\{0\right\})$.  The coupling parameter $g$ is identified with the `counting field' such that 
\beq
  \left.
    \frac{\partial^n \tilde{z}}{\partial (-i g)^n}
  \right|_{g=0} =  \tilde{z}_n= S_\epsilon^{(n)} (\left\{0\right\})
  \label{dzdg}
.
\eeq

This result mirrors that of full counting statistics \cite{naz03,FCS} in which the dephasing of a probe qubit is related to the cumulants of the number of electrons passed through the device, viz. the zero-frequency correlation functions of current fluctuations $\delta I$.
In this context, the relationship between Keldysh-ordered CGF and dephasing was to be expected. However, by proceeding in the above manner, we have obtained explicit expression for the full CGF, as the inverse Laplace transform of $D(z)$ with $l(z)$ as in \eq{LPD}, as well as for the lowest cumulants
from \eq{ztilcomp}.  Obtaining these expressions from standard Keldysh approach would be non-trivial.  The simplicity arises here from the introduction of the two superoperators $O_\epsilon^\pm$, which  correspond to the two branches of the Keldysh contour.

We have talked considerably about the long-time limit of the behaviour of the qubit.  However, the short-time behaviour is also of interest, particularly with an eye to applications in quantum information processing where we definitely want to avoid the heavy dephasing of the long-time limit.  In such circumstances, only the weak-coupling limit is of interest and thus we restrict our short-time discussion to the second-order approximation.  We have $D(z)$ as before, but from truncating \eq{LPD}, we approximate
\beq
  l(z) = - g^2 
  \eww{\phi_0^\mathrm{E} |
    O_\epsilon \Omega_0^\mathrm{E}(z) {\cal Q}^\mathrm{E} O_\epsilon 
  |\phi_0^\mathrm{E} }
  \label{lSHT}
  ,
\eeq
with first-order contribution removed by considering the appropriate rotating frame.
As shown in appendix \ref{appSHT}, we can explicitly perform the Laplace transform of the $D(z)$ arising from \eq{lSHT} and, correct to order $g^2$ and leading term in $t$, the coherence decays as
\beq
  D(t) \sim 1 
  - {\textstyle \frac{1}{2}} g^2  
  \eww{\phi_0^\mathrm{E} |
    O_\epsilon {\cal Q}^\mathrm{E} O_\epsilon 
  |\phi_0^\mathrm{E} }t^2
  .
\eeq
The leading term is proportional to $t^2$, which is to be compared with the linear dependence that would arise from a simple Markovian decay $D(t) \sim 1 - \Gamma_d t$.  Note also that it is the quantity 
$\sqrt{{\textstyle \frac{1}{2}} g^2  
  \eww{\phi_0^\mathrm{E} |
  O_\epsilon {\cal Q}^\mathrm{E} O_\epsilon 
  |\phi_0^\mathrm{E} }}
$, and not $\Gamma_d$, that determines the time-scale of this initial behaviour.

\section{Orthogonal coupling and classical environment \label{SECclasOG}}
We can also effect the separation between system and environment quantities if environment is classical.  For simplicity, we consider just the orthogonal coupling of \eq{OEOG}.  To proceed, we define the system operator
\beq
  F=
  \rb{
  \begin{array}{cccc}
    1 & 1 & 0 & 0\\
    -1 & 1 & 0 & 0 \\
    0 & 0 & 1 & 1 \\
    0 & 0 & -1 & 1
  \end{array}
  }
  ,
\eeq
and use it to transform the effective system Liouvillian of \eq{LeffFULL}.
With ${\cal L}_\mathrm{eff}' =  F{\cal L}^\mathrm{S}_\mathrm{eff} F^{-1}$, we find
\beq
  {\cal L}'_\mathrm{eff}= 
  \rb{
  \begin{array}{cccc}
    0 & 0 & 0 & 0\\
    0 &  {\cal L}'_{22} &0 & {\cal L}'_{24} \\
    0 & 0 & 0 & -i \Delta\\
    0 & {\cal L}'_{42} &  -i \Delta & {\cal L}'_{44}
  \end{array}
  }
  \label{LeffclasOG}
  ,
\eeq
with elements
\beq
  {\cal L}'_{22} &=& -g^2 
    \eww{\phi^\mathrm{E}_0| A O_\epsilon^+ \Omega_+ {\cal Q}^\mathrm{E} O_\epsilon^+|\phi^\mathrm{E}_0}
  \nonumber\\
  {\cal L}'_{24} &=& i  g
    \eww{\phi^\mathrm{E}_0| A  O_\epsilon^+|\phi^\mathrm{E}_0}
  \nonumber\\
  {\cal L}'_{42} &=& i  g
     \eww{\phi^\mathrm{E}_0|B O_\epsilon^+ |\phi^\mathrm{E}_0}
  \nonumber\\
  {\cal L}'_{44} &=& -g^2 
    \eww{\phi^\mathrm{E}_0|B O_\epsilon^+ \Omega_0 {\cal Q} O_\epsilon^+ |\phi^\mathrm{E}_0}
    ,
\eeq
operators
\beq
  A &=& \frac{1}{1+g^2 O_\epsilon^+ \widetilde{\Omega}_+ {\cal Q}^\mathrm{E} O_\epsilon^+ \widetilde{\Omega}_0 {\cal Q}^\mathrm{E}}
  \nonumber\\
  B &=& \frac{1}{1+g^2 O_\epsilon^+ \widetilde{\Omega}_0 {\cal Q}^\mathrm{E} O_\epsilon^+ \widetilde{\Omega}_+ {\cal Q}^\mathrm{E}}
  ,
\eeq
and
\beq
  \widetilde{\Omega}_\pm 
  &=&
  \frac{1}{2}
  \left\{
    \Omega_0^\mathrm{E}(z + i \Delta) \pm \Omega_0^\mathrm{E}(z - i \Delta) 
  \right\}
  \nonumber\\
  \widetilde{\Omega}_0 &=& \Omega_0^\mathrm{E}(z)
  .
\eeq
The stationary state in the original basis is
$\rho_\mathrm{stat} ={\textstyle\frac{1}{2}}\rb{ \op{0}{0} + \op{1}{1}}$, 
the fully incoherent superposition of the two qubit states.  This holds true for any non-pure-dephasing coupling to any classical environment without back-action.

\section{Single Electron Transistor}

As a first example, we consider a charge qubit coupled to a SET environment as discussed in Ref.~\cite{gur08}.
The environment consists of a single level coupled to electron reservoirs that can either be occupied or empty.  The infinite bias limit is assumed such that transport is unidirectional, with electrons entering the level with rate $\Gamma_L$ and leaving with rate $\Gamma_R$.  The only 
pertinent density matrix elements are the two populations of an empty or occupied level.  In this empty/full basis, the Liouvillian of single level reads
\beq
  {\cal L}_0^\mathrm{E} 
  =
  \rb{
  \begin{array}{cc}
    -\Gamma_L & \Gamma_R \\
    \Gamma_L & -\Gamma_R
  \end{array}
  }
  .
\eeq
The stationary state of the environment is $\kett{\phi^\mathrm{E}_0} = \Gamma^{-1} \rb{\Gamma_R, \Gamma_L}^\mathrm{T}$, with $\Gamma= \Gamma_L + \Gamma_R$, and $ {\cal L}_0^\mathrm{E} $ has a single non-zero eigenvalue:  $\lambda_1=-\Gamma$.
The charge qubit is taken to couple to the charge in the SET, and we take the coupling operator to be $\epsilon = \Gamma n_\mathrm{SET}$ with $n_\mathrm{SET}$, the number operator for the single-level, and $\Gamma$ setting the energy scale.  The two coupling superoperators are
\beq
  O_\epsilon^+ =  O_\epsilon^-
  = \Gamma
  \rb{
  \begin{array}{cc}
    0 & 0 \\
    0 & 1 
  \end{array}
  },
\eeq
which are observed to be equal since we have a classical environment.

\begin{figure}[tb]
  \begin{center}
  \psfrag{vt}{$v$}
  \psfrag{Gt}{$\Gamma_R t$}  
  \psfrag{gggg01}{$g=0.1$}
  \psfrag{gggg03}{$g=0.3$}
  \psfrag{gggg05}{$g=0.5$}
  \psfrag{gggg1}{$g=1.0$}
  \psfrag{gggg2}{$g=2.0$}
  \psfrag{r12}{$|\rho_{01}|$}
  \epsfig{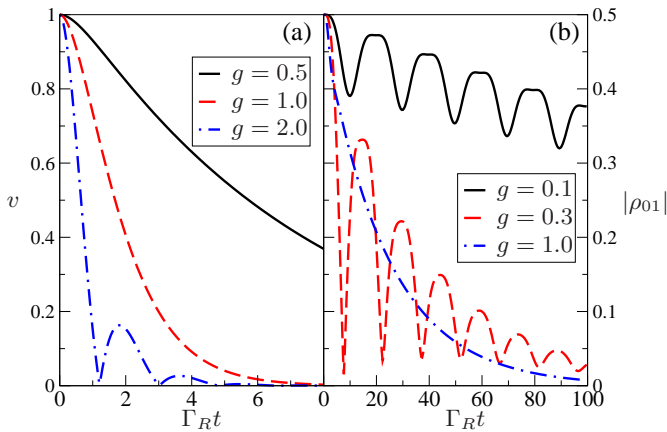}
  \caption{
    Left panel:  The qubit visibility $v(t)$ as a function of time for a SET environment with a pure-dephasing coupling.  As $g$ increases through $g=1$, the behaviour shifts from a simple decay to a highly nonMarkovian oscillating one.
    Right panel: Magnitude of the coherence $|\rho_{01}(t)|$ of the qubit couplied to the SET environment with the orthogonal coupling, starting from an initial state $2^{-1/2}(\ket{0}+\ket{1})$.  Here, the weak-coupling case $g\ll 1$ produces coherence oscillations, whereas strong coupling induces a simple decay.  In all cases, we set $\Gamma_L=\Gamma_R$, and $\Delta=0.3\Gamma_R$.
    \label{SET_PD}
   }
  \end{center}
\end{figure}

With the qubit coupled to this environment in a pure dephasing configuration, the effective Liouvillian term $l(z)$ of \eq{LPD} evaluates
\beq
  l(z) = - i g\frac{\Gamma_L \rb{\Gamma + z}}{ \rb{\Gamma + z} + i g \Gamma_R}
  .
\eeq
The simplicity of this model means that the inverse Laplace transform of $D(z) = (z-l(z))^{-1}$ can be performed analytically.  With $\Gamma_L=\Gamma_R$ for simplicity, we have
\beq
  D(t) = 
  e^{- i g \Gamma_R t} e^{-\Gamma_R t}
  \left\{
    \cosh(\zeta \Gamma_R t) + 
   \zeta^{-1} \sinh(\zeta \Gamma_R t)
  \right\}
  ,
  \nonumber\\
  \label{SETDt}
\eeq
with $\zeta = \sqrt{1-g^2}$.  For weak coupling, $g\ll 1$, this expression correct to second order, becomes
\beq
  D(t) \approx   e^{- i g \Gamma_R t} e^{- \gamma t}
  \left\{
    1 + \frac{\gamma}{2 \Gamma_R}\rb{1-e^{-2(\Gamma_R-\gamma)t}}
  \right\}
  \label{SETDtg2}
  ,
\eeq
with weak-coupling dephasing rate
\beq
  \gamma = \frac{g^2 \Gamma_R}{2}
  .
\eeq
In the strong coupling limit,
 $g \gg 1$,  we obtain
\beq
  D(t) \approx 
  e^{- i g \Gamma_R t}
  e^{-\Gamma_R t}
  \left\{
    \cos(\nu t) + g^{-1} \sin( \nu t)
  \right\}
  ,
\eeq   
with oscillations at a frequency
\beq
  \nu = \rb{g - \frac{1}{2 g}}\Gamma_R,
\eeq
and a decaying envelope of rate $\Gamma_d=\Gamma_R$.
These two contrasting behaviours are illustrated in Fig.~\ref{SET_PD}a in which we plot the visibility\cite{ned07} $v(t) = |D(t)| $.  
Our approach here therefore reproduces the salient features of the pure-dephasing results for a classical telegraph noise model\cite{gal05,ned07,abe08}.

Taking the long-time limit of \eq{SETDt}, we obtain the CGF for the SET charge operator
\beq
  \tilde{z} = \Gamma_R\rb{-i g -1 + \sqrt{1-g^2}}
  .
\eeq
The first few correlation functions are determined from \eq{dzdg} as 
$\Gamma_R\rb{1,0,-3,0,45,\ldots-1575,0,\ldots}$ such that the odd-functions above the first are zero, and are seen to grow with increasing order $k$ as $\sim b^k$.

With the SET coupled to the qubit with a $\sigma_x^\mathrm{S}$ coupling, S and E contributions can be separated along the lines of section \ref{SECclasOG}.  With $ \Gamma_L = \Gamma_R$, the effective Liouvillian in the basis of \eq{LeffclasOG} is 
\beq
  {\cal L}'_\mathrm{eff} = 
  \rb{
  \begin{array}{cccc}
    0 & 0 & 0 & 0\\
    0 & \kappa (z+ 2 \Gamma_R) & 0 & i g \Gamma_R (1+\kappa) \\
    0 & 0 & 0 & - i\Delta \\
    0 & i g \Gamma_R (1+\kappa) &  i\Delta & \kappa\frac{\Delta^2 + (z+ 2 \Gamma_R)^2}{(z+ 2 \Gamma_R)}
  \end{array}
  }
\eeq
with
\beq
  \kappa=-\frac{(g \Gamma_R)^2}{ \Delta^2 + (g  \Gamma_R)^2 + (z+ 2 \Gamma_R)^2}
  .
\eeq
With off-diagonal elements in the effective Liouvillian, the behaviour of the qubit is more complex than in the pure-dephasing case. At first order in $g$, the environment induces a $\sigma_x^\mathrm{S}$ term in the system Liouvillian.  In the weakly coupled case, $g\ll 1$ , this leads to coherent oscillations between the qubit states, as Figure~\ref{SET_PD}b shows.  These oscillations are damped with rates $\Gamma_r = 2g^2  \Gamma_R^3/(\Delta^2 + 4 \Gamma_R^2)$ and $\Gamma_d =\Gamma_r/2$.  As $g$ increases, the damping becomes stronger until the oscillations are suppressed for $g\gg 1,\Delta/\Gamma_R$.
This behaviour with increasing $g$ is exactly the opposite as found for the pure-dephasing model.


\section{Double Quantum Dot environment}
\begin{figure}[tb]
  \begin{center}
  \psfrag{ml}{{\huge $\mu_L$}}
  \psfrag{mr}{{\huge $\mu_R$}}
  \epsfig{file=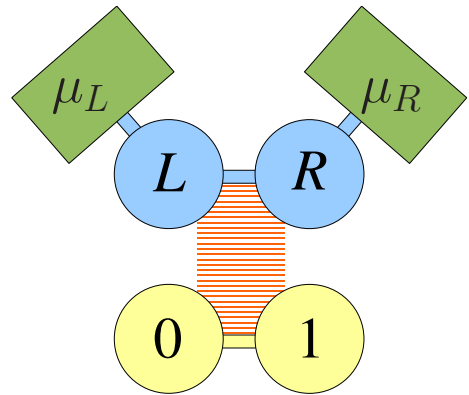, clip=true,width=0.7\linewidth}
  \caption{
    Our second example consists of a charge qubit system (labelled 0 and 1), 
    coupled to environment consisting of a double quantum dot E (labelled $L$ and $R$), and environment E' consists of two electron reservoirs (labelled $\mu_L$ and $\mu_R$).  The DQD is biased such that it is in the nonlinear transport regime.
    \label{FIG_DQD}
   }
  \end{center}
\end{figure}

We now consider a charge qubit coupled to the electronic transport through a DQD
\cite{sto96,ell02,Brandes04,EMAB07}, as depicted in Fig.~\ref{FIG_DQD}.  The crucial difference between this and the previous SET environment is that here the position of the electron in the DQD is a quantum-mechanical variable \cite{hay03,kiess07}, and thus requires a full GME treatment, and not just a rate equation.

We assume the strong Coulomb blockade limit such that the only relevant states are the empty DQD, and a single electron in either left or right dot.  In pseudospin language, the Hamiltonian of an electron in the DQD reads ${\cal H}_\mathrm{E} = \frac{1}{2}\varepsilon \sigma_z^\mathrm{E} + T_c\sigma_x^\mathrm{E}$, with detuning $\varepsilon$ and tunnelling element $T_c$. The two levels are coupled to their respective leads with rates $\Gamma_L$ and $\Gamma_R$, .  The Liouvillian for such a DQD\cite{EMAB07} 
may be written in the basis $\rb{\rho_{00},\rho_{LL},\rho_{RR},\rho_{LR},\rho_{RL}}$ as 
\beq
  {\cal L}^\mathrm{E}_0=
  \rb{
    \begin{array}{ccccc}
       -\Gamma_L & 0 & \Gamma_R & 0 & 0 \\
       \Gamma_L  & 0 & 0 & i T_c & - i T_c\\
       0   & 0 & -\Gamma_R & -i T_c & i T_c \\
       0   & i T_c & -i T_c & -i \varepsilon - {\textstyle\frac{1}{2}}\Gamma_R & 0\\
       0   & -i T_c & i T_c & 0 & i \varepsilon -  {\textstyle\frac{1}{2}}\Gamma_R
    \end{array}
  }
  \label{LDQD}
  .
\eeq
The coherent level splitting is $\Delta_\mathrm{DQD} = \sqrt{\varepsilon^2 + 4T_c^2}$ and in the following we will assume the rates to be equal $\Gamma_L=\Gamma_R$.
We will also assume that the qubit can be coupled to the DQD environment through any of the three Pauli matrices, as well as through $N$, the total occupation of the DQD.  A related model was discussed in Ref.~\cite{lam07} but with transport through both sets of quantum dots, i.e. both through the DQD of the environment and the DQD of our charge qubit.

\begin{figure}[tb]
  \begin{center}
  \psfrag{s1}{$S^{(1)}_\epsilon$}
  \psfrag{s2}{$S^{(2)}_\epsilon$}
  \psfrag{s3}{$S^{(3)}_\epsilon$}
  \psfrag{s4}{$S^{(4)}_\epsilon$}
  \psfrag{eps}{$\varepsilon/\Gamma_R$}
  \psfrag{SX}{$\sigma_x$}
  \psfrag{SY}{$\sigma_y$}
  \psfrag{SZ}{$\sigma_z$}
  \psfrag{Q}{$N$}
  \epsfig{file=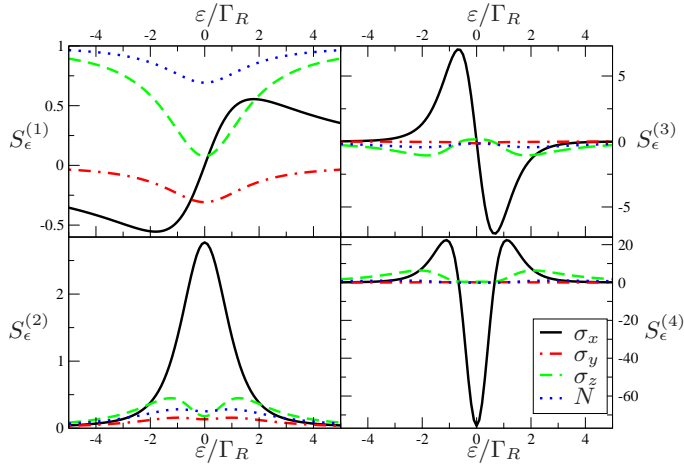, clip=true,width=\linewidth}
  \psfrag{eps}{$\varepsilon$}
  \caption{
    \label{FIG_cums}
    The first four zero-frequency correlation functions $S^{(n)}_\epsilon(0)$, $n=1,\ldots,4$, of operators $\epsilon=\sigma_x$, $\sigma_y$, $\sigma_z$, and $N$ for the DQD environment as a function of DQD detuning $\varepsilon/\Gamma_R$.  The tunnel coupling was $T_c =\Gamma_R$, with $\Gamma_L=\Gamma_R$.
   }
  \end{center}
\end{figure}

Assuming that we can couple to any of these operators,  the qubit in pure-dephasing configuration allows us to perform a complete tomography of the DQD fluctuations through an application of the results of section \ref{SECPD}. By this it is meant that, since the density matrix $\rho^\mathrm{E}$ is completely determined by measurement of the expectation values of these four operators, the long-time fluctuation spectrum is also completely mapped by the correlation functions of the same four operators.
In Figure \ref{FIG_cums}, we plot this set of correlation functions up to fourth order as determined from the generating function $\tilde{z}$.  All correlation functions are even functions of $\varepsilon$ with the exception of the odd-numbered functions for $\sigma_x$.  Operator $\sigma_x$ also stands out as having fluctuations which grow extremely rapidly with increasing order.  
In the long time limit, the expectation value of 
$\sigma_y$ gives the stationary current through the DQD, as expected \cite{sto96}. Surprisingly, although $\sigma_y$ can not in general be identified as the current operator through the DQD, the second correlation function of $\sigma_y$ here does correctly reproduce the DQD shotnoise \cite{ell02,kiess07}.

\begin{figure}[tb]
  \begin{center}
  \psfrag{vt}{$v$}
  \psfrag{Gt}{$\Gamma_R t$}
  \psfrag{gxt}{$\gamma_x t $}
  \epsfig{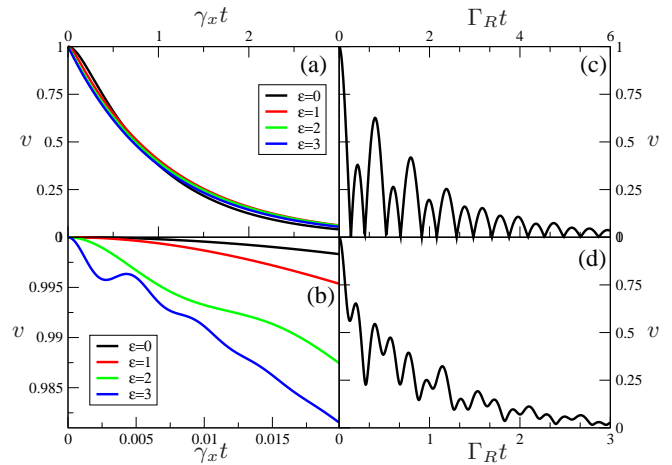}
  \caption{
    The visibility $v(t)$ of the charge qubit coupled to the DQD environment with 
    $\sigma^\mathrm{S}_z \sigma^\mathrm{E}_x$ coupling.
    Lefthand panels: Weak coupling behaviour with $g =0.2$ for different values of $\varepsilon/\Gamma_R$.  The time axes are scaled with weak coupling rate $\gamma_x(\varepsilon)$. 
    On longer timescales (a), the behaviour is approximately Markovian but at short times, deviations are apparent (b).
    Righthand panels: The behaviour for strong coupling is highly nonMarkovian, with pronounced visibility oscillations.
    SE coupling strength and  DQD detuning were here $g=8$, $\varepsilon=0$ (c) 
    and $g=4$, $\varepsilon=2$ (d).  The time axes are scaled with $\Gamma_R$.
    For all plots, we set $T_c =\Gamma_R$.
    \label{FIG_sxtime}
   }
  \end{center}
\end{figure}

We now turn our attention to the dynamics of the qubit itself, and focus first on the pure dephasing case.
The first coupling operator that we consider is
$
  {\cal V}_\mathrm{SE} = \frac{1}{2} \Gamma_R \sigma^\mathrm{S}_z \sigma^\mathrm{E}_x
$.  
In the weak coupling limit,
$g \ll 1$, the dephasing is well described by the expansion of \eq{LPD} up to second order.
Figure \ref{FIG_sxtime} shows the behaviour of the qubit visibility over both long and short time scales.  The dephasing rate $\Gamma_d=\gamma_x(\varepsilon)$ is easily calculated from \eq{G2}, but its form is too unwieldy to give here.
In the special case of $\varepsilon=0$, however, it reads
$
  \gamma_x = \gamma_x(\varepsilon=0)=
    2 g^2 \Gamma_R (\Gamma_R^2 + 8 T_c^2)
    /
    (
      \Gamma_R + 12  T_c^2/\Gamma_R
    )
$.
The Corresponding visibility as a function of time is
\beq
  v_{\varepsilon=0}(t) = \frac{\Gamma_R + 2 \gamma_x}{\Gamma_R} e^{- \gamma_x t} 
  - \frac{2 \gamma_x}{\Gamma_R} e^{-(\Gamma_R/2 - \gamma_x)t} 
  \label{vtPDx}
  .
\eeq
which, at short times, reads $v(t) = 1 - \gamma_x \Gamma_R t^2/2$, showing the generic deviation from Markovian behaviour.
For finite $\varepsilon$, the expression for the visibility is more complicated and, as Fig.~\ref{FIG_sxtime} shows, describes oscillations at short times. This nonMarkovian effect is due to the coherent coupling of the system to the environment, with the oscillations proceeding with a frequency $\sim\Delta_\mathrm{DQD}$.  It should be noted that these oscillations will only be visible if the DQD dynamics are not too heavily damped \cite{EMAB07}.

The $\varepsilon=0$ case is a particularly special limit of this model in which
the environment is, at second-order in $g$, identical to the SET, as a comparison of \eq{vtPDx} and \eq{SETDtg2} shows.  The reason for this equivalence is most easily seen by rewriting \eq{LDQD} in the basis $\left(\rho_{00},
\rho_{LL},\rho_{RR}, \mathrm{Re}(\rho_{LR}),
\mathrm{Im}(\rho_{LR})\right)$: 
\beq
  \rb{
  \begin{array}{ccccc}
    -\Gamma_L & 0 & \Gamma_R & 0 &0\\
    \Gamma_L & 0 & 0 & 0 & - 2 T_c \\
    0 & 0 & -\Gamma_R & 0 & 2 T_c \\
    0 & 0 & 0 & -\frac{1}{2}\Gamma_R &  \epsilon\\
    0 & T_c & - T_c &  - \epsilon &  - \frac{1}{2}\Gamma_R
  \end{array}
  }
  . \nonumber
\eeq
It is then clear that for $\epsilon=0$, the real part of the interdot coherence,
$\mathrm{Re}(\rho_{LR})$, decouples from the rest of the (free) environment dynamics.
At second order, the coupling operator $\sigma_x^\mathrm{DQD}$ couples only to this real part of the coherence and the DQD environment therefore appears as a classical environment with a single excited level of width $\Gamma_R/2$.
At orders higher than second, this equivalence breaks down as the operator $\sigma_x^\mathrm{E}$ couples to further parts of the environment density matrix.

For stronger coupling $g \gtrsim 1$, nonMarkovian effects dominate.  We consider first the case with $\varepsilon=0$.  For $g\sim 1 $ the visibility behaves like that of the strongly-coupled SET with oscillations at frequency $\sim g \Gamma_R$ and an envelope that decays as $\Gamma_R/2$.
Increasing  $g$ further leads to a doubling in the period of the oscillations not seen in the SET model.  This behaviour can be traced to  the increased importance of a further pair of poles of $D(z)$ oscillating at a frequency $\sim 2 T_c$ for $g\gg 1$.
As illustrated by Fig. \ref{FIG_sxtime}d, the situation for $\epsilon \ne 0$ is more complicated, with frequencies associated with all coherent couplings of the SE complex playing a role.

\begin{figure}[tb]
  \begin{center}
  \psfrag{vt}{$v$}
  \psfrag{Gt}{$\Gamma_R t$}
  \psfrag{rhoL}{$\rho_\mathrm{LL}$}
 \epsfig{file=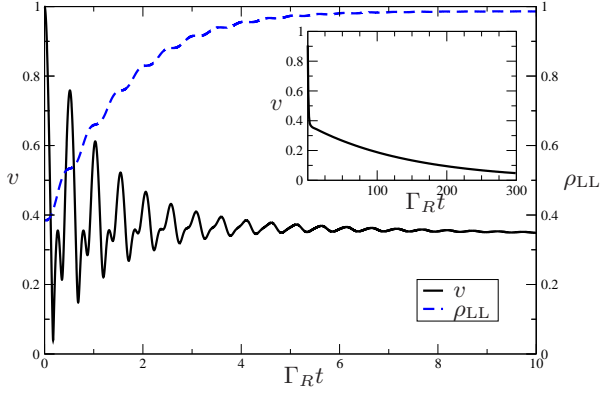, clip=true,width=0.9\linewidth}
  \caption{
    The visibility $v(t)$ as a function of time for the coupling 
    ${\cal V}_\mathrm{SE} = \frac{1}{2} \Gamma_R \sigma^\mathrm{S}_z \sigma^\mathrm{E}_z$.  The main panel shows that the visibility decays with an initial rate $\Gamma_R/2$ until it reaches a value which appears stationary on the scale of this plot.  Also shown is the population of the left dot of the  DQD, $\rho_{\mathrm{LL}}$, which shows localisation due to the Zeno effect induced by the coupling to the qubit.
    The inset shows the visibility on a much longer time-scale, which reveals a slow decay with rate $\gamma_\mathrm{Zeno}$.
    The parameters were $g=12$, $T_c = \Gamma_R$ and $\varepsilon=0$.
    \label{FIG_zeno}
   }
  \end{center}
\end{figure}

Now consider a second SE coupling operator:
$
  {\cal V}_\mathrm{SE} = \frac{1}{2} \Gamma_R \sigma^\mathrm{S}_z \sigma^\mathrm{E}_z
$.
For weak coupling, the behaviour is qualitatively similar to that for the previous $\sigma^\mathrm{E}_x$-coupling, with the $\varepsilon=0$ dephasing rate given by
$
  \gamma_z= 4 g^2 \Gamma_R^3 T_c^2 (13 \Gamma_R^2 + 36 T_c^2)/(\Gamma_R^2 + 12 T_c^2)^3
$.
The strong coupling behaviour is, however, significantly different, as can be seen from Fig.~\ref{FIG_zeno}.  Initially, the visibility shows oscillations and decays with rate $\Gamma_R/2$. But instead of the visibility decaying to zero after a
few times $\Gamma_R^{-1}$, it reaches a value which, on the time scale of the initial decay, appears constant.
This is a manifestation of quantum Zeno effect, with the coupling to $\sigma^\mathrm{E}_z$ effectively measuring and localising an environmental electron in the lefthand dot of the DQD. Transport through the DQD, and along with it the dephasing of the qubit, is then almost completely inhibited.
For large but finite coupling, this localisation is not perfect, but decays with the rate
$
  \gamma_\mathrm{Zeno} = 
  T_c^2/(g^2 \Gamma_R)
$ for $\varepsilon=0$.
Needless to say, this effect relies on having a quantum environment.

\begin{figure}[t]
  \begin{center}  
  \psfrag{vt}{$v(t)$}
  \psfrag{Gt}{$\Gamma_R t$}
  \psfrag{r}{$\rho$}
  \psfrag{r00}{$\rho_{00}$}
  \psfrag{r11}{$\rho_{11}$}
  \psfrag{eps}{$\varepsilon/\Gamma_R$}
  \psfrag{sz}{$\ew{\sigma^\mathrm{S}_z}$}
  
  \epsfig{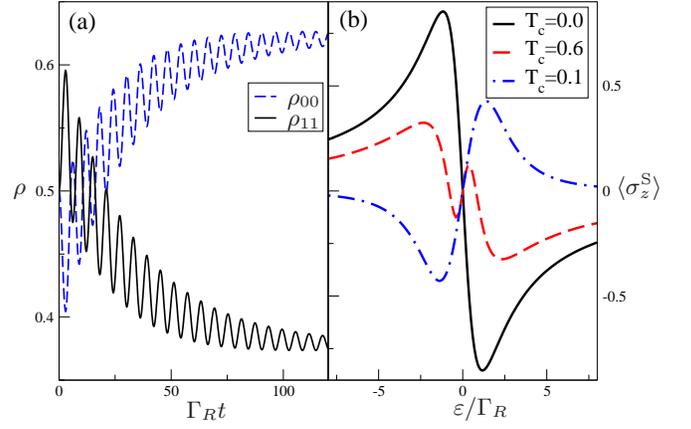}
  \caption{
   DQD model with  $\sigma^\mathrm{S}_x \sigma^\mathrm{E}_x$ coupling.
   Left: Population of the two qubit states as a function of time starting from an equal superposition with $\varepsilon = \Gamma_R/2$, $g=0.4$  and $\Delta=T_c=\Gamma_R$.  A population difference develops due to the quantum nature of the environment.
   Right: The stationary qubit population difference $\ew{\sigma_z^\mathrm{S}}$ as a function of $\varepsilon$ for various values of $T_c/\Gamma_R$.  For $\varepsilon=0$, we have $\ew{\sigma_z^\mathrm{S}}=0$ and the relaxation of the qubit is effectively classical.  Away from this point, quantum back-action leads to a finite value of $\ew{\sigma_z^\mathrm{S}}$.
   Parameters were $g=0.2$, and  $\Delta=\Gamma_R$.
    \label{FIG_nonclas}
   }
  \end{center}
\end{figure}

Finally, we consider couplings to the qubit such that it undergoes not only dephasing but also relaxation. The behaviour is correspondingly more complex and here we focus on just one aspect: the effects of quantum back-action.
In section \ref{SECclasOG} we saw that for a qubit coupled with any non-pure-dephasing coupling
to a classical environment, the stationary qubit state is always $\rho_\mathrm{stat}^\mathrm{S}={\textstyle{\frac{1}{2}}}\rb{\op{0}{0}+\op{1}{1}}$.
Let us therefore consider a SE coupling 
$
  {\cal V}_\mathrm{SE} = \frac{1}{2} \Gamma_R \sigma^\mathrm{S}_x \sigma^\mathrm{E}_x
$ and focus on the weak-coupling limit.
Fig.~\ref{FIG_nonclas} shows the behaviour of the populations of the two qubit levels as a function of time for $\varepsilon=\Gamma_R/2$.  As is clear, a steady state is reached in which $\rho_{00}\ne\rho_{11}$, or in other words for which $\ew{\sigma_z}$ is in the steady state is non-zero.
Due to the dephasing of off-diagonal elements,  the purity of the final state is sinply given by $P=\ew{\sigma_z}^2$. In Fig.~\ref{FIG_nonclas}b we plot  $\ew{\sigma_z}$ as a function of detuning, from which it is  immediately clear that $\ew{\sigma_z}=0$ only if $\varepsilon=0$ and that  $\ew{\sigma_z}$ is antisymmetric as a functions of $\varepsilon$.

This behaviour is evidence of the back-action of the system on the environment. 
For $\varepsilon = 0$, we effectively have a classical model in the weak-coupling limit, as discussed above, and thus $\ew{\sigma_z}=0$ follows accordingly.  Even for large couplings, this model always relaxes to the completely mixed state for $\varepsilon = 0$, implying that all coupling between system and environment are essentially classical and back-action free.
For $\varepsilon \ne 0$, the coupling to the qubit disturbs the motion of the environment which in turn feeds back into the behaviour of the qubit and leads to a finite value of $\ew{\sigma_z}$ in the stationary limit.

The relevance of the foregoing results for experimental quantum dot systems can be assessed as follows. The weak-coupling Markovian dephasing/relaxation rates for the models considered here are of the order $g^2 \Gamma$, with $\Gamma$ some typical environment rate and where, in principle, $g$ can be as small as one wishes. Finite detector bandwidth should therefore not be a problem, provided sufficiently a weak SE coupling can be realised.  
Conversely, the nonMarkovian effects occur on energy scales of either $\Gamma$ or at the Bohr frequencies of the environment $\Delta^\mathrm{E}$.
In early DQD experiments, such as Ref.~\cite{hay03}, the tunnelling rates and coherent coupling strength were of the order of $\sim$10MHz.  Observing the nonMarkovian effects produced by an environment with such time-scales would be very challenging.
However, more recent experiments \cite{gus06,fri07} operate with coupling parameters in the range 1-10kHz.  At such frequencies, the passage of a single electrons through a quantum dot can be detected. In a set-up with such parameter, we would expect the nonMarkovian effects described here to lie within the experimentally observable range.

\section{summary}

In summary, we have presented a formalism for investigating the dynamics of quantum systems coupled to nonequillibrium environments that exhibit a full range of fluctuation statistics. The end product of this analysis is an effective system Liouvillian which describes the systems behaviour without need to explicitly follow the environmental dynamics.

With a weak-coupling expansion, this effective Liouvillian can be expressed in terms of environmental correlation functions, showing how nonGaussian fluctuations impact the system behaviour.  This represents a systematisation and generalisation of calculations of Ref.~\cite{gur08} to arbitrary systems and environments, as well as to environmental fluctuations beyond Gaussian.
Away from the weak coupling limit, two situations distinguish themselves where the dependence of the systems behaviour on environmental correlation functions of all orders can be made explicit ---
pure dephasing coupling and when the environment is classical. 
The pure dephasing case is interesting as here the qubit can be thought of as a detector of the environment fluctuations --- a measurement of the dephasing of the qubit delivers a generating function for the statistics of environmental operator through with the system and environment are coupled.

We have considered two example environments from mesoscopic transport ---the SET and DQD, acting on
a charge qubit.  These examples serve to illustrate the differences between 
quantum and classical environments, the latter class expanded to include situations, such as we saw with the zero-detuning DQD with $\sigma^\mathrm{E}_x$-coupling, where a quantum environment acts as if it were classical.
With a pure dephasing coupling, weak coupling results always show deviations from Markovian behaviour at short times, but only the quantum environment has the capacity to induce oscillations in the visibility.  At strong couplings, the behaviour of qubit highly non-Markovian, with the behaviour in a classical environment quite simple, but that in a quantum environment far more complex with phenomena such as the quantum Zeno effect occurring.
Finally, with a coupling that induces relaxation, quantum environments distinguish themselves from classical ones, by inducing a steady state of the qubit system with a non-zero purity.  This is attributed to the quantum back-action of the system of the environment.

\appendix
\section{Third-order correlation functions \label{appO3}}
\begin{widetext}

The third-order contribution to the effective system Liouvillian is expressed in \eq{L3} in terms of the following two correlation functions
\beq
   \bar{S}_\epsilon^{(3a)}(z, z') &=& 
    \eww{
        O^+_\epsilon \Omega^\mathrm{E}(z){\cal Q}
        O^+_\epsilon \Omega^\mathrm{E}(z'){\cal Q}
        O^+_\epsilon
    }
   \nonumber\\
   &=& 
     \int_{0}^\infty d(t_3-t_2) \int_{0}^\infty d(t_2-t_1)
     e^{-z (t_3-t_2) -z'(t_2-t_1) }
     \left\{
       \ew{\delta\epsilon(t_3)\delta\epsilon(t_2)\delta\epsilon(t_1)}
       -\ew{\epsilon}\ew{\delta\epsilon(t_3)\delta\epsilon(t_1)}
     \right\}
   \nonumber\\
   \bar{S}_\epsilon^{(3b)}(z, z') &=& 
   \eww{
      O^+_\epsilon \Omega^\mathrm{E}(z){\cal Q}
      O^+_\epsilon \Omega^\mathrm{E}(z'){\cal Q}
      O^-_\epsilon
   }
    \nonumber\\
   &=&
   \int_{0}^\infty d(t_3-t_2) \int_{0}^\infty d(t_2-t_1)
     e^{-z (t_3-t_2) -z'(t_2-t_1) }     
     \left\{
       \ew{\delta\epsilon(t_1)\delta\epsilon(t_3)\delta\epsilon(t_2)}
        -\ew{\epsilon}\ew{\delta\epsilon(t_1)\delta\epsilon(t_3)}
     \right\}
    \nonumber\\
\eeq
In the long time limit, the generating function $\widetilde{z}$ of section \ref{SECPD} is comprised of Keldysh-ordered correlation functions. The third-order correlator\cite{sal06} is
\beq
  S_{\cal K}^{(3)} (\omega_2,\omega_1)
  =
  \int_{-\infty}^\infty d(t_3-t_2) \int_{-\infty}^\infty d(t_2-t_1)
  e^{i \omega_1(t_2-t_1) + i \omega_2 (t_3-t_2)}
  S_{\cal K}^{(3)} (t_3,t_2,t_1).
\eeq
with
\beq
  S_{\cal K}^{(3)} (t_3,t_2,t_1) &=&  
  \rb{\frac{1}{2}}^3
  {\cal T_K}
  \ew{   
    (\delta\epsilon_+ (t_3)+\delta\epsilon_- (t_3))
    (\delta\epsilon_+ (t_2)+\delta\epsilon_- (t_2))
    (\delta\epsilon_+ (t_1)+\delta\epsilon_- (t_1))
  },
\eeq
where $ {\cal T_K}$ is the Keldysh-ordering symbol and subscripts $\pm$ correspond to forward and backward branches of the Keldysh contour.
\end{widetext}

\section{Short time, weak coupling\label{appSHT}}
With a pure-dephasing coupling, the coherence factor correct to second order in $g$ is
\beq
  D(z) = \frac{1}{z-g^2 \sum_{k\ne0}\frac{c_k}{z-\lambda_k}}
\eeq
with
\beq
  c_k =   
  \eww{\phi_0^\mathrm{E} | O_\epsilon |\phi_k^\mathrm{E} }
  \eww{\phi_k^\mathrm{E} | O_\epsilon |\phi_0^\mathrm{E} }
  .
\eeq
and $  O_\epsilon =\frac{1}{2}\rb{O_\epsilon^+ +   O_\epsilon^-}$.
Introducing $\alpha_k = g^2 c_k / \lambda_k^2$ and writing the dephasing rate of \eq{G2} explicitly as 
$
  \Gamma_d = - g^2  \sum_{k\ne0} c_k/\lambda_k = - \sum_{k\ne0}  \alpha_k\lambda_k
$,
$D(z)$ can be written in a partial fraction decomposition as
\beq
  D(z) = \rb{1 +\sum_{k\ne0}  \alpha_k } \frac{1}{z+\Gamma_d}
  - \sum_{k\ne0} \frac{ \alpha_k }{z-\lambda_k+\alpha_k}
\eeq
It is then trivial to perform the inverse Laplace transform to obtain
\beq
  D(t) = \rb{1 +\sum_{k\ne0}  \alpha_k } e^{-\Gamma_dt}
  - \sum_{k\ne0} \alpha_k e^{(\lambda_k-\alpha_k)t}
\eeq
A short time thus expansion yields
\beq
  D(t) &\sim& 1 -{\textstyle \frac{1}{2}}g^2 t^2 \sum_{k\ne0}  c_k
  \nonumber\\
  &=&
  1 
  - {\textstyle \frac{1}{2}} g^2  
  \eww{\phi_0^\mathrm{E} |
    O_\epsilon {\cal Q}^\mathrm{E} O_\epsilon 
  |\phi_0^\mathrm{E} }t^2
  ,
\eeq
correct to second order in $g$ and leading term in $t$.
\section*{Acknowledgements}
Work supported by the WE Heraeus foundation and by DFG grant BR 1528/5-1. I am grateful to S.~A.~Gurvitz for providing the inspiration to study this problem, and to T.~Brandes for discussions.


\end{document}